\documentclass[prb,showpacs,twocolumn]{revtex4}

\usepackage{amsmath,amssymb}

\newcommand{\bu}[1]{{#1}^1B_u}
\newcommand{\but}[1]{{#1}^3B_u}
\newcommand{\buminus}[1]{{#1}^1B_u^-}
\newcommand{\ag}[1]{{#1}^1A_g}
\newcommand{\agtr}[1]{{#1}^3A_g}
\newcommand{\agplus}[1]{{#1}^1A_g^+}

\usepackage{graphicx}

\begin{document}

\widetext

\title{Computational Investigations of the Primary Excited States of Poly(\textit{para}-phenylene vinylene)}

\author{Robert J. Bursill$^1$ and William Barford$^{2}$}

\affiliation{
$^1$School of Physics, University of New South Wales, Sydney, New South Wales 2052, Australia\\
$^2$Department of Physics and Astronomy, University of Sheffield, Sheffield, S3 7RH, United Kingdom
}

\begin{abstract}
The Pariser-Parr-Pople model of $\pi$-conjugated electrons is
solved by the density matrix renormalization group method for the
light emitting polymer, poly(\textit{para}-phenylene vinylene).
The energies of the primary excited states are calculated. When
solid state screening is incorporated into the model parameters there
is excellent agreement between theory and experiment, enabling
an identification of the origin of the key spectroscopic features.
\end{abstract}


\maketitle

\section{Introduction}\label{Se:1}

Since the discovery of electroluminescence in poly(para-phenylene
vinylene) (PPV) in 1990\cite{burroughes90}, a  variety of
experimental and theoretical techniques have been deployed to
investigate the physics of the primary photoexcitations of the
phenyl-based light emitting polymers. Linear and non-linear
optical spectroscopies have revealed the energies and symmetries
of the dominant dipole-allowed transitions, as well as the
important dipole-forbidden transitions that participate in
non-linear optical processes.

All phenyl-based light emitting polymers exhibit a characteristic
absorption spectra. These show two dominant  excitations polarized
parallel to the chain axis (at $2.8$ eV and $6.1$ eV in
PPV\cite{martin99}). There are also two intermediate weaker
transitions (at $3.6$ eV and $4.8$ eV in PPV\cite{martin99}). The
higher of these intermediate transitions is predominantly
polarized perpendicular to the chain axis, whereas the
polarization of the lower transition is less well-defined. (In PPV
it is predominately polarized parallel to the
long-axis\cite{miller}.) Furthermore, the strength of this lower
transition is enhanced by chemical substitution.
Electroabsorption\cite{martin99} and two-photon
absorption\cite{frolov02} reveal a dipole-forbidden state at
approximately $0.7$ eV above the lowest dipole-allowed transition.
These two states are usually labelled the $\ag{m}$ and $\bu{1}$
states, respectively. Finally, a triplet state has been observed
at $0.7$ eV below the $\bu{1}$ excitation (see ref\cite{kohler04}
for references), with another dipole connected triplet state $1.4$
eV higher in energy\cite{monkman01}. (Notice that this higher
triplet state, labelled as the $\agtr{m}$ state, is virtually
degenerate with its singlet counterpart, namely the $\ag{m}$
state. These two states are often referred to as the singlet and
triplet charge-transfer states. Their degeneracy can be explained
by the fact that they are the lowest pseudomomentum branches of
the $n=2$ Mott-Wannier excitons, which have odd parity
electron-hole wavefunctions and therefore experience no exchange
interactions\cite{barford02}.)

An important early insight into the nature of the primary
photoexcitations of the phenyl-based systems was  provided by Rice
and Gartstein\cite{rice94,gartstein95}, who argued that they can
essentially be understood as arising from the delocalization of
the primitive benzene excitations. Kirova and Brazovskii, on the
other hand, have argued that a conventional semiconductor band
picture of bound particle-hole excitations is a more appropriate
description\cite{kirova99}. Other theoretical work on PPV
includes, a single configuration interaction (CI) calculation of
the Pariser-Parr-Pople model by Chandross and
Mazumdar\cite{chandross97}; density matrix renormalization group
calculations on reduced molecular-orbital models by Barford
\emph{et al.}\cite{barford97}; quantum chemistry calculations on
the INDO Hamiltonian by Beljonne \emph{et al.}\cite{beljonne99},
and Weibel and Yaron\cite{weibel02}; and \emph{ab initio}
Bethe-Salpeter equation (BSE) calculations by Rohlfing and
Louie\cite{rohlfing99}. These theoretical predictions will be
discussed more fully in Section \ref{Se:3}, when we discuss the
results of the calculations presented here and their comparisons
to experiment.

In this paper we present density matrix renormalization group
(DMRG) calculations on the full Pariser-Parr-Pople  model of PPV.
The DMRG method for calculating the full electronic spectra of
conjugated polymers has a number of advantages over its
competitors. It is more accurate than single or double CI
calculations, and does not suffer from finite-size consistency
errors; it makes no assumptions about the nature of the
excitations (unlike the BSE method, which assumes that the
excitations are particle-hole pairs), thus it is able to
accurately model highly correlated excited states; and unlike
standard quantum chemistry techniques, it is able to calculate the
excited spectra of large molecules. The DMRG method is most
readily suited to reduced basis Hamiltonians, such as the
$\pi$-electron Pariser-Parr-Pople model. This is not necessarily a
disadvantage over \emph{ab initio} methods, as when correctly
parameterized the Pariser-Parr-Pople model makes very accurate
predictions\cite{bursill98, chandross97}. Moreover, the
Pariser-Parr-Pople model possess particle-hole symmetry, which
means that spatial, particle-hole and spin-flip symmetries can be
employed to target high-lying excited states in the DMRG method
(see ref\cite{barford05}, for example).

DMRG calculations on the full Pariser-Parr-Pople model were
presented for poly(\emph{para}-phenylene) by Bursill  and
Barford\cite{bursill02}. This paper extends that approach to PPV.
In the next section we define the Pariser-Parr-Pople model and
briefly describe our implementation of the DMRG method. In Section
\ref{Se:3} we describe and discuss our results, and
compare them to other approaches, concluding in section
\ref{Se:4}.

\section{Model and Methodology}\label{Se:2}

\subsection{The Pariser-Parr-Pople Model}

The Pariser-Parr-Pople  model is a $\pi$-electron model of conjugated polymers, defined by,
\begin{eqnarray}\label{Eq:1}
{\cal H}
& = &
- \sum_{<ij>\,\sigma} t_{ij}
    \left[ c_{i\sigma}^{\dagger} c_{j\sigma} + c_{j\sigma}^{\dagger} c_{i\sigma} \right]
\nonumber
\\
& &
+\; U \sum_{i}
    \left(n_{i\uparrow}-1/2\right)
    \left(n_{i\downarrow}-1/2\right)
\nonumber
\\
& &
+\;
\frac{1}{2} \sum_{i\neq j} V_{ij} (n_i - 1)(n_j - 1),
\end{eqnarray}
where $<>$ represents nearest neighbors, $c_{i\sigma}$ destroys a $\pi$-electron on site $i$, $n_{i\sigma} =
c_{i\sigma}^{\dagger} c_{i \sigma}$, and $n_i = n_{i\uparrow} +
n_{i\downarrow}$.

We use the Ohno parameterization for the Coulomb
interaction, defined by,
\begin{equation}\label{Eq:2}
     V_{ij} = U / \sqrt{ 1 + ( U \epsilon r_{ij}/14.397)^2 },
\end{equation}
where $r_{ij}$ is the inter-atomic distance (in \AA), $U$ is the
on-site Coulomb interaction (in eV), and $\epsilon$ is the
dielectric  constant. This interaction is an interpolation between
an on-site Coulomb repulsion, $U$, and a Coulomb potential,
$\textrm{e}^2/4\pi\epsilon\epsilon_0 r_{ij}$ as $r_{ij}
\rightarrow \infty$.

\begin{figure}[tb]
\begin{center}
\includegraphics[scale=0.50]{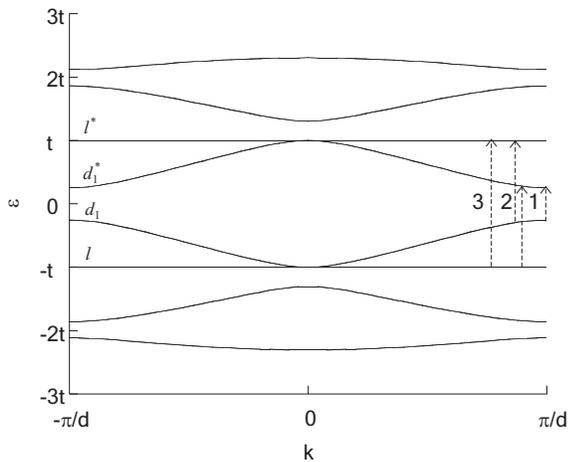}
\end{center}
\caption{The band structure of PPV with all nearest neighbor bond integrals, $t$,
equal. The low-lying particle-hole transitions are labelled $1$, $2$
and $3$.} \label{Fi:1}
\end{figure}

The band structure, obtained in the non-interacting limit ($U=0$),
is shown in Fig.\ \ref{Fi:1}. The pair of  non-bonding bands are a
consequence of the $D_{2h}$ symmetry of the Hamiltonian in the
non-interacting limit with only nearest-neighbor bond integrals.
The low-lying particle-hole excitations are labelled $1$ (for
transitions involving $d_1$ and $d_1^*$), the degenerate pair
labelled $2$ (for transitions involving $d_1$ and $l^*$, and $l$
and $d_1^*$), and $3$ (for transitions involving $l$ and $l^*$).
Coulomb interactions lift the degeneracy of the intermediate pair
(which become the intermediate pair of transitions described in
Section \ref{Se:1}), while the transitions $1$ and $3$ become the
low and high energy transitions described in Section \ref{Se:1}.
Evidently, to obtain a more realistic description of the excited states it is
necessary to include Coulomb interactions.

In the following we use two sets of parameters to model the
excited states. The first set, called the \emph{optimized}
parameters, were derived by fitting the predicted
Pariser-Parr-Pople excitation energies of stilbene to the
experimental spectrum of stilbene in vacuo\cite{castleton99}.
These are, $t_p = 2.539$ eV, $t_d = 2.684$ eV, $t_s = 2.22$ eV, $U
= 10.06$ eV, and $\epsilon = 1$ (where the phenyl, double and
single bond integrals are defined in Fig.\ \ref{Fi:2}). Although
giving instructive results, this parameter set fails to
quantitatively predict the excitation energies of polymers in the
solid state, as they fail to account for the solvation effects of
the surrounding dielectric\cite{moore98, barford04}. The other
parameter set, called the \emph{screened} parameters, were derived
by Chandross and Mazumdar\cite{chandross97} to account for solid
state solvation effects. These parameters are, $t_p = 2.4$ eV,
$t_d = 2.6$ eV, $t_s = 2.2$ eV, $U = 8$ eV, and $\epsilon = 2$.

\subsection{The Density Matrix Renormalization Group (DMRG) Method}

Eq.\ (\ref{Eq:1}) is solved by the DMRG method for chains of up to
$28$ phenyl rings (i.e.\ $222$ sites). The DMRG method is an
efficient truncation procedure for solving quantum lattice
Hamiltonians, especially in one-dimension\cite{white}. The details
of the DMRG implementation for this problem, particularly the
procedure of deriving an optimized reduced basis for the phenyl
ring\cite{zhang, bursill02, barford02b}, are described in detail
in ref\cite{bursill02}. Also shown in ref\cite{bursill02} are comprehensive DMRG convergence tests that are applicable to phenyl-based systems.

As described in Section \ref{Se:1}, the Pariser-Parr-Pople model
possess spin-flip and particle-hole symmetries, which  we exploit
in the DMRG method to target excited states. With only onsite and
nearest neighbor Coulomb interactions the Pariser-Parr-Pople model
applied to the PPV structure formally possess $D_{2h}$ symmetry.
In this limit this means that the eigenstates are labelled with
the spatial symmetry assignments, $A_g$, $B_{1u}$, $B_{2u}$, and
$B_{3g}$. Longer range Coulomb interactions with the PPV structure
reduce the $D_{2h}$ symmetry to $C_{2}$ symmetry, and thus the
true spatial symmetry assignments are $A_g$ and $B_{u}$. However,
it is computationally expedient, with very little loss of
accuracy, to retain $D_{2h}$ symmetry while incorporating long
range Coulomb interactions. This is achieved by taking the bond
angle between the single and double bonds in the vinylene unit to
be $180^0$, rather than $120^0$. To ensure that the overall
molecular size is consistent to the original structure the single
and double lengths are reduced to $1.283$ \AA~ and $1.194$ \AA,
respectively. The phenyl bond length is $1.40$ \AA. The next
section describes the calculated results.

\section{Results and Discussions}\label{Se:3}

\subsection{DMRG Calculations}

\begin{figure}[tb]
\begin{center}
\includegraphics[scale=0.50]{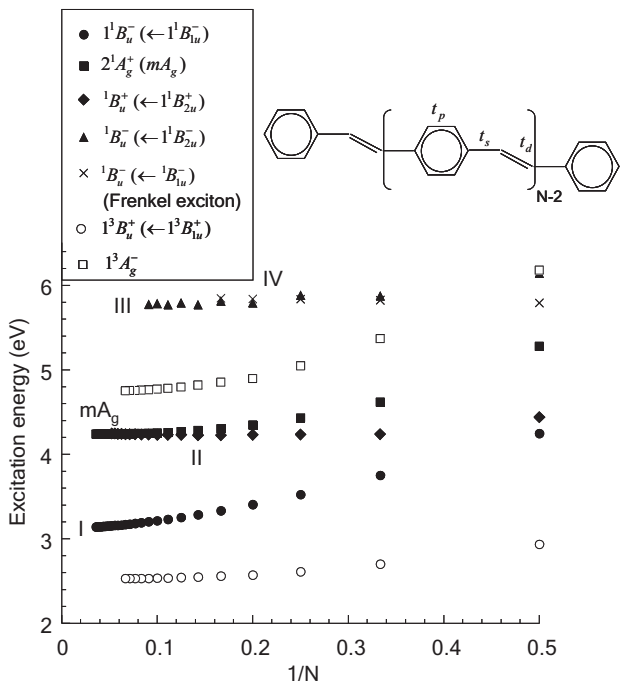}
\end{center}
\caption{The DMRG calculated transition energies of
\textit{para}-phenylene vinylene oligomers as a function of
inverse chain length, calculated from the Pariser-Parr-Pople model
with optimized parameters: $U=10.06$ eV, $t_p = 2.539$ eV, $t_d =
2.684$ eV, $t_s = 2.22$ eV, and dielectric constant, $\epsilon =
1$. The excited states are identified with the spectroscopic
features shown in Fig.\ 3 of ref\protect\cite{martin99} and Fig.\
1 of ref\protect\cite{frolov02}. The symmetry assignments of the
states shown in brackets are the symmetries the eigenstates would
have if PPV had $D_{2h}$ rather than $C_2$ symmetry. The inset
shows the oligo-phenylene vinylene structure with the bond
integrals $t_p$, $t_d$ and $t_s$. } \label{Fi:2}
\end{figure}

Fig.\ \ref{Fi:2} shows the DMRG calculated excitation energies of
oligo(\textit{para}-phenylene vinylenes)  using the
Pariser-Parr-Pople model with unscreened parameters ($U=10.06$ eV,
$t_p = 2.539$ eV, $t_d = 2.684$ eV, $t_s = 2.22$ eV, and $\epsilon
= 1$). ($N=2$ corresponds to stilbene.) The $\buminus{1}$ state is
the strong lowest energy dipole allowed transition (labelled I in
Fig.\ 3 of ref\cite{martin99}). The strong transition dipole
moment between this state and the $\agplus{2}$ state indicates
that the $\agplus{2}$ is the state labelled `$\ag{m}$' in
non-linear optical spectroscopies (see, for example, Fig.\ 1(a) in
ref\cite{frolov02}). The energies of the $\buminus{1}$ and
$\agplus{2}$ states initially reduce rapidly as a function of
chain length, indicating that these states reduce their kinetic
energy by readily delocalizing the exciton wavefunction along the
chain. This is primarily because these states are largely
constructed from particle-hole transitions between the bonding and
anti-bonding bands, labelled $d_1$ and $d_1^*$ in Fig.\
\ref{Fi:1}\cite{footnote1}. A description of these states as
strongly bound band exciton states is therefore appropriate in the
long-chain limit. Indeed, the $\buminus{1}$ state is the lowest
pseudomomentum branch of the $n=1$ Mott-Wannier exciton, while the
$\agplus{2}$ is the lowest pseudomomentum branch of the $n=2$
Mott-Wannier exciton\cite{barford02,barford05}. (Higher lying
pseudo-momentum branches of the $n=1$ and $n=2$ excitons for DMRG
calculations on poly(para-phenylene) are illusrated in Fig.\ 5 of
ref\cite{bursill02}.) Notice that this assignment places a
\emph{lower bound} on the binding energy of the $n=1$ exciton as
$E(mA_g) - E(1B_u)$.

We next consider the state labelled II and shown by diamonds in
Fig.\ \ref{Fi:2}. This state has $^1B_{u}^+$ symmetry, but would
have $^1B_{2u}^+$  symmetry if PPV had $D_{2h}$ symmetry. From a
band picture analysis, this state is composed of an antisymmetric
combination of particle-hole transitions from $d_1$ to $l^*$ and
$l$ to $d_1^*$. At the Pariser-Parr-Pople model level, it has
positive particle-hole symmetry, and is therefore
dipole-forbidden. However, it is weakly dipole allowed  in real
systems, and its oscillator strength is further enhanced by
chemical substitution. Its excitation energy is virtually
independent of chain length, because, first its wavefunction is
composed of non-bonding orbitals, and therefore there is no
hybridization via one-electron transfer terms, and second since it
has a very small oscillator strength, resonant exciton transfer
along the chain is virtually inoperative. These two features (weak
oscillator strength and a chain independent energy) strongly
indicate that the second spectroscopic feature (labelled II in
Fig.\ 3 of ref\cite{martin99}) originates from this state.

The state labelled III and shown by triangles in Fig.\ \ref{Fi:2}
has $^1B_{u}^-$ symmetry, but would have $^1B_{2u}^-$ symmetry if
PPV had $D_{2h}$ symmetry. From a band picture analysis, this
state is composed of a symmetric combination of particle-hole
transitions from $d_1$ to $l^*$ and $l$ to $d_1^*$. It has
negative particle-hole symmetry, and therefore has an allowed
dipole transition from the ground state. This state is the third
spectroscopic feature of PPV (labelled III in Fig.\ 3 of
ref\cite{martin99}).

The state labelled IV and shown by crosses in Fig.\ \ref{Fi:2} has
$^1B_{u}^-$ symmetry, but would have $^1B_{1u}^-$ symmetry if PPV
had $D_{2h}$ symmetry. From a band picture analysis, this state is
composed of  particle-hole transitions from $l$ to $l^*$. It has negative particle-hole symmetry, has a strong
dipole transition from the ground state, and is polarized along the
chain axis. This state is the fourth spectroscopic feature of PPV
(labelled IV in Fig.\ 3 of ref\cite{martin99}), as is usually
referred to as the `Frenkel' exciton, as its particle-hole
wavefunction is typically confined to a single phenyl-ring.

Finally, the two triplet states, $1^3B_u^+$ and $m^3A_g^-$, are
also shown in Fig.\ \ref{Fi:2}.

The results shown in Fig.\ \ref{Fi:2} are obtained from the
Pariser-Parr-Pople model with optimized parameters that are not
parametrized to model solid state solvation effects. The
dielectric response of the environment is predicted to red-shift
the intra-molecular excitations of a single chain by up to $0.1$
eV for the $1^1B_u$ state, $0.7$ eV for the $m^1A_g$ state, and
$1.5$ eV for the charge gap\cite{moore98,barford04}. We would
therefore not expect these calculations on a single chain to
compare directly with experimental observations.

\begin{figure}[tb]
\begin{center}
\includegraphics[scale=0.50]{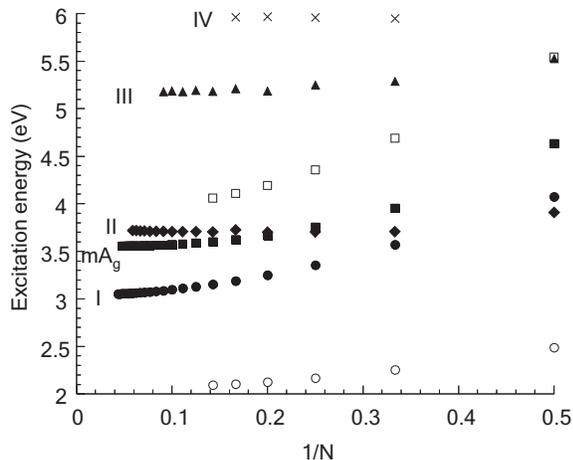}
\end{center}
\caption{The DMRG calculated transition energies of
\textit{para}-phenylene vinylene oligomers as a function of
inverse chain length, calculated from the Pariser-Parr-Pople model
with screened parameters: $U=8$ eV, $t_p = 2.4$ eV, $t_d = 2.6$
eV, $t_s = 2.2$ eV, and $\epsilon = 2$. The symbols are defined in
the inset of Fig.\ \ref{Fi:2}.  The excited states are identified
with the spectroscopic features shown in Fig.\ 3 of
ref\protect\cite{martin99} and Fig.\ 1 of
ref\protect\cite{frolov02}.} \label{Fi:3}
\end{figure}

Chandross and Mazumdar attempted to model solvation affects by
adjusting the Pariser-Parr-Pople parameters and introducing a
static dielectric constant\cite{chandross97}. Although the
introduction of a \emph{static} dielectric constant is not a
faithful representation of the dielectric response (because the
timescale for the particle-hole motion is very similar to the
timescale of the dielectric response\cite{moore98}), these
screened parameters do lead to Pariser-Parr-Pople model
predictions that are remarkably close to the experimental
observations.

Fig.\ \ref{Fi:3} shows the DMRG calculated results with the
screened parameters ($U=8$ eV, $t_p = 2.4$ eV, $t_d = 2.6$ eV,
$t_s = 2.2$ eV, and $\epsilon = 2$). Comparing this figure with
Fig.\ \ref{Fi:2} we see that the $\buminus{1}$ and $\agplus{2}$
states are red-shifted by ca.\ $0.1$ eV and $0.6$ eV,
respectively, as expected from solvation effects. The excited
states are identified with the spectroscopic features shown in
Fig.\ 3 of ref\protect\cite{martin99}. Their energies are
remarkably consistent with the experimental values of $2.8$,
$3.6$, $4.8$, and $6.1$ eV obtained in ref\cite{martin99}, lending
additional credence to our assignments of the excited state
origins of the features, discussed above.

\subsection{Other Approaches}

We now compare our DMRG predictions of the Pariser-Par-Pople model
with other approaches. Our results using the screened parameters
are consistent with those of Chandross and Mazumdar\cite{chandross97}. They
used SCI on the Pariser-Par-Pople model. For an eight-unit
oligomer they  calculate the $\bu{1}$ state at $2.7$ eV, an
$m^1A_g$ state at $3.3$ eV, and the $\bu{n}$ state at 3.6 eV. The
$\bu{1}$ and $\ag{m}$ states are the $n=1$ and $n=2$ excitons,
while the $\bu{n}$ state coincides with the charge-gap and
therefore indicates the onset of the particle-hole continuum. They
also predict the $\but{1}$ state at $1.4$ eV.

An \textit{ab initio} BSE calculation by Rohlfing and
Louie\cite{rohlfing99} on a PPV polymer predicts dipole allowed
and forbidden singlet excitons at $2.4$ eV and $2.8$ eV,
respectively, with the quasi-particle gap at $3.3$ eV. They also
predict triplet excitons at $1.5$ eV and $2.7$ eV. The $2.4$ eV
and $2.8$ eV singlet excitons are the $1^1B_u$ and $m^1A_g$
states, respectively, while the $1.5$ eV and $2.7$ eV triplet
excitons are the $1^3B_u$ and $m^3A_g$ states, respectively. The
$m^1A_g$ and $m^3A_g$ states are nearly degenerate, as predicted
by the Mott-Wannier exciton theory for odd parity particle-hole
wavefunctions. Using the same technique with a screened
electron-hole interaction van der Horst \textit{et
al.}\cite{horst01}  predict a $1^1B_u$ binding energy in PPV of
$0.48$ eV.

The origin of the higher-lying transitions has also been
investigated. Rohfling and Louie\cite{rohlfing99},  and Weibel and
Yaron\cite{weibel02} predict that peak II in PPV arises from an
exciton caused predominately by the antisymmetric combination  of
particle-hole transitions from $d_1$ to $l^*$ and $l$ to $d_1^*$.
Weibel and Yaron\cite{weibel02} have also investigated the effects
of breaking particle-hole symmetry on the oscillator strength and
polarization of peak II. Using the semiempirical INDO Hamiltonian
on nonplanar di-hydroxy-PPV, their calculations indicate that
chemical substitution and mixing of the $\pi$ and $\sigma$
orbitals enhances the oscillator strength, as originally suggested
by Gartstein \emph{et al.}\cite{gartstein95} Moreover, as
illustrated in Fig.\ 5 of ref\cite{weibel02}, this peak becomes
predominately polarized along the chain axis, in agreement with
experiment\cite{miller}.

\section{Conclusions}\label{Se:4}

Gathering together the various theoretical predictions of the
origins of the primary excited states  of PPV we now interpret the
key spectroscopic features of PPV.

Peak I corresponds to the low-energy dipole active $1^1B_u^-$
state. This is  the lowest pseudomomentum branch of the family of
$n=1$ Mott-Wannier singlet excitons resulting from the Coulomb
attraction between the particle-hole excitation from the valence
($d_1$) to the conduction ($d_1^*$) bands. Approximately $0.7$ eV
higher in energy is the $m^1A_g$ state, identified by
electroabsorption\cite{martin99}, two-photon absorption and
photoinduced  absorption\cite{frolov02}. The Pariser-Parr-Pople
model calculations described in this paper suggest that this state
is the $2^1A_g^+$ state, which is
 the lowest pseudomomentum branch of the family of
$n=2$ Mott-Wannier excitons.  This assignment places a lower bound
on the  spectroscopically determined  binding energy of the
$\bu{1}$ exciton of  $0.7$ eV. Approximately $0.7$ eV below the
$1^1B_u^-$ exciton is the $1^3B_u^+$ triplet, indicating a large
exchange energy characteristic of correlated states. This state is
the lowest pseudomomentum branch of the family of $n=1$
Mott-Wannier triplet excitons. Photo-induced absorption from the
$1^3B_u^+$ triplet indicates another triplet, the $1^3A_g^-$
state, at approximately $1.4$ eV higher in energy, and essentially
degenerate with the $2^1A_g^+$ state. This triplet state is   the
lowest pseudomomentum branch of the family of $n=2$ Mott-Wannier
triplet excitons. As expected from Mott-Wannier exciton theory in
one-dimension\cite{barford02, barford05}, the odd  particle-hole
parity singlet and triplet (charge-transfer) excitons are
virtually degenerate. The $\bu{n}$ state at $0.1$ eV higher in
energy than the $\ag{m}$ state in PPV\cite{martin99} indicates
binding energies of $\sim 0.8$ eV and $0.1$ eV for the $n=1$ and
$n=2$ singlet excitons, respectively. Higher in energy are the
excitations associated with peaks II and III. These arise from
antisymmetric and symmetric combinations  of particle-hole
transitions from $d_1$ to $l^*$ and $l$ to $d_1^*$, respectively.
Finally, peak IV is the intraphenyl, or Frenkel, exciton. This
identification of the primary excited states of PPV also applies
to other phenyl-based systems, e.g.\
PPP\cite{bursill02,barford05}.

The calculations presented in this paper have entirely concerned
vertical transitions, where the excited  states are calculated in
the geometry of the ground state. Of course, electron-lattice
relaxation is an important process in polymers that determines
Stokes shifts, self-trapping, and exciton and charge transfer
between polymers. The consequences of electron-lattice relaxation
on the excited states energies and structures have been considered
in refs\cite{el}.

\begin{acknowledgements}
R.\ J.\ B.\ acknowledges support from the Australian Research Council and the J.\ G.\ Russell Foundation.
W.\ B.\  thanks  the Leverhulme Trust for financial support.
\end{acknowledgements}

\end{document}